Short Paper

# Perceived Importance of ICT Proficiency for Teaching, Learning, and Career Progression among Physical Education Teachers in Pampanga

Kristine Joy D. Magallanes
Mexico Campus, Don Honorio Ventura State University, Philippines
kristinejoy.d.magallanes@gmail.com
(corresponding author)

Mark Brianne C. Carreon
Mexico Campus, Don Honorio Ventura State University, Philippines

Kristalyn C. Miclat
Mexico Campus, Don Honorio Ventura State University, Philippines

Niña Vina V. Salita
Mexico Campus, Don Honorio Ventura State University, Philippines

Gino A. Sumilhig
Mexico Campus, Don Honorio Ventura State University, Philippines

Raymart Christopher C. Guevarra
Mexico Campus, Don Honorio Ventura State University, Philippines

John Paul P. Miranda
Mexico Campus, Don Honorio Ventura State University, Philippines
jppmiranda@dhvsu.edu.ph

**Abstract**

The integration of information and communication technology (ICT) has become increasingly vital across various educational fields, including physical education (PE). This study aimed to evaluate the proficiency levels of PE teachers in using various ICT applications and to examine the relationship between the perceived importance of ICT proficiency for teaching and learning, career advancement, and actual proficiency among Senior High school PE teachers in the municipality of Mexico, Pampanga. This study employed a quantitative descriptive approach. PE teachers from the municipality of Mexico, Pampanga, were selected as the respondents. This study used a two-part survey. The first section collected demographic data, such as age, gender, rank/position, and years of teaching experience, and the second section assessed ICT skill levels and the perceived importance of ICT in teaching, learning, and career progression. The results revealed that the majority of PE teachers had access to ICT resources. However, their proficiency levels with these tools varied significantly. Factors such as age, teaching experience, and professional position were found to significantly influence teachers' proficiency and their perceptions of the benefits of ICT integration in PE instruction. The study provided a glimpse of the current state of ICT integration among Senior High school PE teachers in Mexico, Pampanga, Philippines. This also highlights areas of improvement. The study suggests that policymakers, administrators, and training program developers should focus on enhancing the ICT proficiency of PE teachers to improve teaching practices and student engagement. Enhancing the ICT proficiency of PE teachers is recommended to foster better teaching experiences, increase student engagement, and promote overall educational outcomes.

**INTRODUCTION**

In recent decades, the technological revolution has brought about significant changes that have influenced modern societies (Wastiau et al., 2013; Kretschman, 2015). Information and Communication Technology (ICT) has become an integral part of our daily lives, jobs, and education systems. Today, ICT is recognized as a critical consideration for educational establishments worldwide (Abed & Mohamed, 2018). In the educational sector, ICT plays a key role by creating new demands and changes that significantly affect teachers, generating a constant need for training and updating (Martínez-Rico et al., 2022). For teachers and students, ICT enhances the teaching process and student learning. This involves more than just integrating ICT into lessons; it allows teachers to effectively utilize school equipment, especially in PE, to enhance their daily teaching practices.

The integration of ICT in education is essential for effective teaching and learning, particularly for PE teachers who can leverage technology to improve instructional methods and student engagement. Recognizing the perceived importance of ICT proficiency for teaching, learning, and career progression among PE teachers is vital for developing targeted professional development programs and institutional support mechanisms. This study aims to explore these perceptions among Physical Education teachers in Pampanga, highlighting the necessity for continuous ICT training and its impact on their professional growth. Moreover, a cross-sectional study design was employed to measure the exposure of a population at a single point in time and place (Simkus, 2023). This type of study allows researchers to select respondents based on specific inclusion and exclusion criteria, measure their exposure to ICT, and examine the outcomes (Setia, 2016). By studying the associations between these variables, the research aims to provide insights into the current state of ICT proficiency among PE teachers and its implications for their teaching, learning, and career progression.

## *Conceptual Framework*

The model (Figure 1) shows the factors affecting the perceived importance of ICT proficiency for PE teachers on the following premises: learning, teaching, and career progression. The relationship between demographic factors such as age, rank, and years of teaching PE in Mexico and Pampanga has an influence on school teachers' proficiency in using and integrating ICT in their learning, teaching, and career progression.

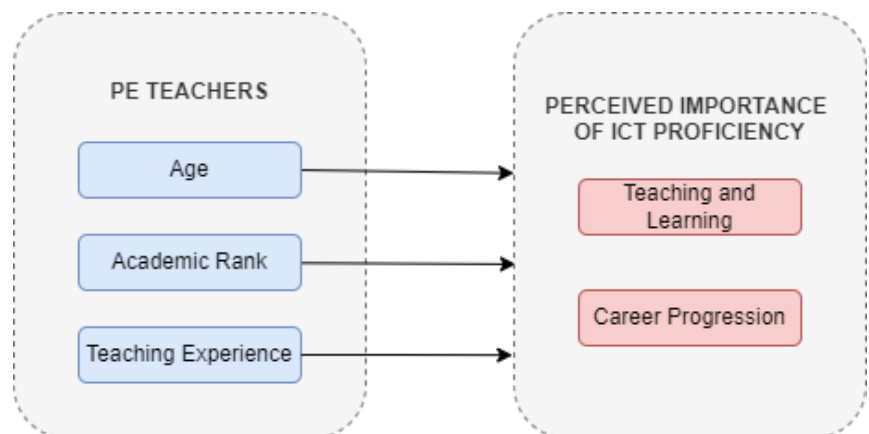

*Figure 1.* Conceptual framework

## *Research Problems*

This study aims to explore PE teachers' evaluations of different ICT applications and ascertain whether these perceptions differ significantly depending on local demographic factors. This study specifically sought to answer the following questions:

1. What is the level of proficiency of the respondents in using ICT applications?
2. How do respondents prioritize the importance of proficiency in different ICT applications for their teaching and learning, and career progression?
3. What is the association between the perceived importance of ICT proficiency for teaching and learning, career progression, and proficiency in various applications?

## LITERATURE REVIEW

### *Integration of ICT in Education*

Teaching effectiveness involves activities by teachers that promote students' cognitive, affective, and psychosocial development. ICT media, such as videos, digital cameras, and internet resources, are instrumental in conveying information on various physical activities like aerobic dance, cycling, aquatics, and athletics, particularly in PE. UNESCO (2023) emphasized the need for professional development opportunities for teachers to enhance their use of ICT in formative assessments, individualized instruction, online resources, and student interaction. ICT offers flexible and effective methods for teachers to improve their competencies, maintain their roles, and connect with the global teacher community. However, without proper support, teachers may limit their use of ICT to skill-based applications, thereby restricting students' academic growth. It is vital for education managers, supervisors, teacher educators, and decision-makers to be trained in ICT to support teachers in adopting innovative teaching methods.

Furthermore, ICT has the potential to significantly enhance teaching practices across various subjects, including PE. By leveraging ICT tools and resources, educators can improve learning experiences, increase engagement, facilitate problem-solving, enhance communication, develop research skills, and improve decision-making processes (Adarkwah, 2020). The integration of ICT in education is essential for equipping students with crucial 21st-century skills such as information literacy, media literacy, and ICT literacy (Aslan & Zhu, 2016). Additionally, ICT enhances the quality of instruction, teaching effectiveness, resource accessibility, and management techniques in educational settings (Dyantyi, 2023). In PE, ICT can introduce innovative instructional approaches, such as interactive simulations, video analysis software, fitness tracking apps, and online resources, which enhance students' learning experiences and engagement (Batanero et al., 2019). ICT training for PE teachers, especially for addressing the needs of students with disabilities, fosters educational innovation and inclusivity (Batanero et al., 2019).

The effective integration of ICT in teaching practices benefits both students and teachers. Professional development programs and a supportive organizational culture are crucial in encouraging teachers to adopt ICT-based practices (Soomro, 2023). The use of ICT in teaching subjects like Physics has been shown to positively impact the learning process. This shows the importance of employing appropriate ICT tools to effectively convey knowledge (Hussaini, 2023). Despite these, the integration of ICT in education presents significant opportunities for enhancing teaching and learning, it also requires targeted professional development, institutional support, and a commitment to ongoing innovation to fully realize its potential.

## *Obstacles to Integrating ICT in Education*

ICT in education faces several obstacles despite widespread enthusiasm among educators. Research highlights a significant lack of subject-specific professional development, leading to limited access to ICT resources in school PE departments. Several factors contribute to teachers' struggles with ICT integration, including insufficient time for training, lack of confidence and motivation, entrenched teaching habits, and fear of change (Hew & Brush, 2007; Levin & Wadmany, 2008; Player-Koro, 2012; Gilakjani, 2013; Kollia et al., 2020). Personal characteristics also play a role, with variables such as educational level, age, gender, experience, and attitude towards computers influencing ICT adoption. In addition, less experienced teachers often exhibit higher ICT competence, yet these factors collectively create substantial barriers to effective ICT integration (Dube et al., 2018). Addressing these challenges requires targeted professional development and a better understanding of teachers' needs to facilitate the effective use of ICT in education.

## METHODOLOGY

### *Research Design*

This study employed a cross-sectional research design which incorporates a descriptive and quantitative approach. Descriptive research outlines the characteristics of the population or phenomena under study, focusing on the "what" rather than the "why" (Manjunatha, 2019). This approach was chosen to provide a thorough overview of current features and trends as it highlights the key patterns without delving into causal explanations. The quantitative method was used to enable statistical analysis to ensure objective measurement and comparison of variables. This combination allows for efficient data collection and provides a detailed snapshot of the phenomena, offering valuable insights for future research and practice.

### *Respondents*

The respondents to this study were physical education instructors working in public senior high schools in and around Mexico, Pampanga, Philippines. Convenience sampling was employed as the sampling method. Moreover, convenience sampling is a non-probability sampling technique in which respondents were chosen based on their availability and desire to participate (Hassan, 2024). In other words, the sample consisted of people that the researcher could easily reach and agreed to participate in the study. The respondents were 19 people, with 10 female and nine male respondents, aged between 24 and 43 years. The academic rank of the respondents was composed of five Teacher I, 10 Teacher II, two Teacher III, one Master Teacher II, and one Special Science Teacher III with two to 18 years of teaching experience (Table 1).

Table 1. Demographic and Professional Details of Respondents

| Category | Details |
|---|---|
| Total Number of Respondents | 19 |
| Sex Distribution | 10 Female, 9 Male |
| Age Range | 24 to 43 years |
| Academic Rank | 5 Teacher I |
| | 10 Teacher II |
| | 2 Teacher III |
| | 1 Master Teacher II |
| | 1 Special Science Teacher III |
| Teaching Experience | 2 to 18 years |

## Instrument

The researchers in this study employed a Likert-scale questionnaire as a research tool. The researchers adapted a questionnaire from two research studies to make it more suitable for this study, wherein the instrument's reliability was ensured through testing again during the investigation, and the Cronbach's alpha was 0.89, indicating the instrument's strong reliability. For validation, two professors who holds doctorate degree in educational management and specializes in educational technology were consulted. Both were consulted to ensure that the instrument is align with the research problems. In the study by Harerimana and Mtshali (2019), a questionnaire was used to measure the skill level of nursing students in using various ICT applications such as email, word processing software, and electronic health records, which are important in nursing education.

In the study by Suominen et al. (2021), a questionnaire was used to investigate the self-assessment of basic ICT skills among new vocational school students in Finland. Both studies used a Likert-scale questionnaire to assess respondents' ICT skills. The newly constructed questionnaire was composed of parts 1 and 2, wherein part 1 focused on the PE teachers' demographic profile, such as name (optional), age, rank, and years of teaching. The second part of the questionnaire comprised two sets of scales. The first part was a 5-point scale ranging from 1 = novice, 2 - intermediate, 3 = competent, 4 = advanced, and 5 = expert, focusing on the proficiency of PE teachers in ICT (Table 2). The second part of the scale is a 4-point scale that rates the most important to least important of ICT in terms of teaching, learning, and career progression.

Table 2. Ranges, Descriptions, and Interpretations

| Rating | Range | Description | Interpretation |
|---|---|---|---|
| 1 | 1.00 – 1.80 | Novice | Needs significant improvement |
| 2 | 1.81 – 2.60 | Intermediate | Basic understanding/ skill |
| 3 | 2.61 – 3.40 | Competent | Good understanding/ skill |
| 4 | 3.41 – 4.20 | Advanced | Very good/ excellent skill |
| 5 | 4.21 – 5.00 | Expert | Outstanding/ exceptional skill |

## RESULTS AND DISCUSSION

### *Level of Proficiency in ICT Applications*

**Word Processing Application**

For the word processing application (WPA), users rated the ease of performing tasks, such as creating a new word document (Mean = 4.68, SD = 0.82), saving a document onto the computer (Mean = 4.68, SD = 0.82), opening an existing document (Mean = 4.58, SD = 0.838), changing font size and style (Mean = 4.63, SD = 0.831), inserting pictures (Mean = 4.63, SD = 0.831), inserting clip art (Mean = 4.58, SD = 0.838), inserting charts (Mean = 4.32, SD = 0.885), inserting tables (Mean = 4.63, SD = 0.831), splitting texts into columns (Mean = 4.58, SD = 0.838), and changing page layout (Mean = 4.63, SD = 0.831). The respondents reported an overall mean score of 4.594 with a standard deviation of 0.8363 (Table 3). Effective use of computers requires a solid understanding of word processing. E-learning has been created to enable the widespread distribution of core knowledge and abilities, even outside the classroom (Gupta, 2021).

**Spreadsheet Application**

For spreadsheet application (SA), users evaluated tasks such as saving and opening workbooks (Mean = 4.63, SD = 0.831), adding and deleting worksheets (Mean = 4.53, SD = 0.905), applying background color (Mean = 4.53, SD = 0.772), adding or deleting columns and rows (Mean = 4.42, SD = 0.961), printing data (Mean = 4.53, SD = 0.841), merging and unmerging cells (Mean = 4.42, SD = 0.902), creating tables and graphs (SA7; Mean = 4.26, SD = 0.991), sorting data (Mean = 4.16, SD = 0.958), conversion of formulas to values (Mean = 4.00, SD = 1.155), and using formulas for statistical analysis (SA10; Mean = 3.63, SD = 1.116). The overall mean SA score was 4.311, with a standard deviation of 0.9432 (Table 3). Spreadsheet software promotes the development of computational thinking across various mathematical areas such as arithmetic, equations, trigonometry, and statistics (Borkulo et al., 2018; Borkulo et al. 2023).

**Presentation Application**

For presentation applications (PRE), users assessed tasks such as opening and creating new presentations (Mean = 4.63, SD = 0.831); using templates (Mean = 4.58, SD = 0.838); changing the font format (Mean = 4.63, SD = 0.831); manipulating slides (Mean = 4.53, SD = 0.841); inserting audio, video, and pictures (Mean = 4.47, SD = 0.905); inserting graphs and diagrams (Mean = 4.32, SD = 0.946); inserting links from other websites (Mean = 4.26, SD = 1.098); applying animations and transitions (Mean = 4.47, SD = 0.841); inserting slides from other presentations (Mean = 4.58, SD = 0.838); and exporting presentations to other formats (Mean = 4.42, SD = 0.902). The respondents' overall mean PRE score was 4.489, with a standard deviation of 0.8871 (Table 3). Students learn better if the course material is presented using visual tools. Teachers believe that PowerPoint presentations make the content more appealing, helping capture students' attention (Lari, 2014).

*Conferencing Application*

For conferencing application (CA), users evaluated tasks such as creating and sharing meeting links (Mean = 4.16, SD = 1.119), joining meetings (Mean = 4.58, SD = 0.769), accepting participants (Mean = 4.47, SD = 0.841), using filters (Mean = 4.21, SD = 0.976), muting or unmuting participants (Mean = 4.63, SD = 0.684), chatting with participants (Mean = 4.53, SD = 0.841), changing tiled layout (Mean = 4.26, SD = 0.872), using Jamboard (Mean = 3.74, SD = 0.933), recording meetings (Mean = 4.21, SD = 0.855), and sharing screens (Mean = 4.37, SD = 0.895). The overall mean CA score was 4.316, with a standard deviation of 0.8785 (Table 3). Visual conferencing technology enables the real-time transmission and reception of audio and visual data across a network, allowing users in different locations to hold face-to-face meetings (Kristòf, 2020).

*Photo and Video Editing Application*

For photo and video editing application (PVEA), users provided ratings for tasks such as cropping photos (Mean = 4.53, SD = 0.841), creating picture collages (Mean = 4.53, SD = 0.772), changing video backgrounds (Mean = 4.05, SD = 1.129), removing picture backgrounds (Mean = 4.16, SD = 1.119), making photo prints (Mean = 4.00, SD = 1.155), enhancing photo attributes (Mean = 4.32, SD = 0.946), applying transitions and effects (Mean = 4.16, SD = 1.068), inserting voice-overs in videos (Mean = 3.58, SD = 1.170), applying sound effects (Mean = 3.84, SD = 1.119), and merging video clips (Mean = 3.74, SD = 1.147). The respondents reported an overall mean score of 4.091 with a standard deviation of 1.0466 (Table 3). Videos appeal widely because of their visual and aural components, allowing viewers to uniquely process information. Videos in teaching and learning benefit students, teachers, their institutions, and the entire educational system (NextThought, 2024). In addition, 97% of education experts believe that video is crucial to students' academic experiences (Bennet, 2021).

*Basic Computer Troubleshooting*

For basic computer troubleshooting (BTS), users rated tasks such as installing output devices (Mean = 4.42, SD = 0.769), fixing blue screen errors (Mean = 3.53, SD = 1.219), restarting the computer (Mean = 4.58, SD = 0.692), freeing up hard disk space (Mean = 3.79, SD = 1.182), uninstalling unnecessary programs (Mean = 4.37, SD = 0.955), adjusting Windows visual effects (Mean = 4.05, SD = 0.970), removing viruses and malware (Mean = 3.47, SD = 1.124), resetting internet browsers (Mean = 3.89, SD = 1.243), running system restore (Mean = 3.37, SD = 1.116), and updating software versions (Mean = 3.84, SD = 1.015). The overall mean score for the BTS was 3.931, with a standard deviation of 1.0285 (Table 3). Troubleshooting involves identifying errors, diagnosing faults, and restoring the systems. With the growing complexity and prevalence of automated systems, the ability to troubleshoot is essential for ensuring safety and security in various situations (Bordewieck & Elson, 2021).

Table 3. Overall self-reported level of proficiency

|      | Mean  | SD     | Proficiency | Interpretation                |
|------|-------|--------|-------------|-------------------------------|
| WPA  | 4.594 | 0.8363 | Expert      | Outstanding/ exceptional skill |
| SA   | 4.311 | 0.9432 | Expert      | Outstanding/ exceptional skill |
| PRE  | 4.489 | 0.8871 | Expert      | Outstanding/ exceptional skill |
| CA   | 4.316 | 0.8785 | Expert      | Outstanding/ exceptional skill |
| PVEA | 4.091 | 1.0466 | Advanced    | Very good/ excellent skill    |
| BTS  | 3.931 | 1.0285 | Competent   | Good understanding/skill      |

## *Importance of Proficiency for Teaching and Learning, and Career Progression*

The respondents rank the importance of proficiency in ICT in teaching and learning (LST) such as providing a module using a word processing application (Mean = 3.42, SD = 0.69), creating an assessment or evaluation using the application in assessing student learning (Mean = 3.58, SD = 0.61), creating a classroom schedule and outlining activities using a spreadsheet (Mean = 3.68, SD = 0.58), using spreadsheet in assessing skills, movement patterns, or routines of the students in physical education (3.37, SD = 0.68), using presentation for giving instructions (Mean = 3.74, SD = 0.45), using video presentations in teaching basic skills and steps in physical education (Mean = 3.63, SD = 0.50), creating or editing an instructional video and audio (Mean = 3.58, SD = 0.61), cropping and editing and instructional photo for discussion (Mean = 3.42, SD = 0.69), using output devices in presenting an instructional video or audio (e.g. projector or speaker) (Mean = 3.84, SD = 0.38), creating a video conference in an online setup (Mean = 3.42, SD = 0.51), using a word document in creating test and exam papers (Mean = 3.84, SD = 0.38), using a word document in creating lesson plan (Mean = 3.89, SD = 0.32), using spreadsheets for creating grades (Mean = 3.89, SD = 0.32), using spreadsheets for inputting information of the students (Mean = 3.89, SD = 0.32), creating a presentation in providing course material (Mean = 3.89, SD = 0.32), creating a presentation for teaching a subject (Mean = 3.79, SD = 0.42), creating and editing videos for teaching (Mean = 3.68, SD = 0.48), creating e-portfolios (Mean = 3.58, SD = 0.61), using Google Meet and Zoom in online class (Mean = 3.63, SD = 0.50), and lastly, using a projector in presenting PPT in class (Mean = 3.47, SD = 0.77).

The overall mean score for Learning and Teaching was 3.66, with a standard deviation of 0.33. The integration of computers in education has revolutionized teaching and learning by streamlining activities, delivering accurate data rapidly, enhancing efficiency, and enabling precise task execution through the utilization of operating systems and application packages such as Microsoft Office Word, Excel, and PowerPoint (Kabiru et al., 2021).

In Career Progression (CP), the respondents ranked the importance of ICT, such as I believe I need ICT in my future career (Mean = 3.79, SD = 0.42), I believe that ICT can improve my teaching career (Mean = 3.79, SD = 0.42), I believe that the use of ICT improves my quality of teaching (Mean = 3.83, SD = 0.38), integrating technology increases my teaching productivity (Mean = 3.79, SD = 0.42), and I believe that ICT fits into my teaching skills (Mean = 3.68, SD = 0.58), I believe that ICT makes it easier for me to do my instruction (Mean = 3.79, SD = 0.42), and lastly, I believe that technology is completely compatible with all aspects of my academic work (Mean = 3.68, SD = 0.58). The overall mean score for career progression is 3.77, with a standard deviation of 0.43 ICT can provide more flexible and effective ways or methods for the professional development of teachers to maintain their jobs, improve their competencies, and connect them to the global teacher community (Sahito & Vaisanen, 2017).

### Associations between Perceived Importance of ICT Proficiency for Teaching and Learning, Career Progression, and Proficiency in Various Applications

Table 4 presents the Spearman's association results. This revealed the associations between LST and proficiency in various applications. There was a positive association between LST and WPA proficiency, although it did not reach statistical significance ($r_s$ = 0.411, $p$ = 0.080). A substantial positive association ($r_s$ = 0.664, $p$ = 0.002) was found between LST and SA proficiency, demonstrating a moderate link between the perceived relevance of ICT proficiency in teaching and learning and proficiency in SA. LST and PRE proficiency were found to be substantially associated ($r_s$ = 0.442, $p$ = 0.058), indicating a moderate relationship between the perceived importance of ICT proficiency in teaching and learning and PRE proficiency. There was also a significant positive association between LST and CA proficiency ($r_s$ = 0.625, $p$ = 0.004), indicating a moderate relationship between the perceived importance of ICT proficiency for teaching and learning and CA proficiency.

Furthermore, the results indicated a significant positive connection between LST and PVEA proficiency ($r_s$ = 0.583, $p$ = 0.009), suggesting a significant association between PVEA proficiency and the perceived importance of ICT proficiency in teaching and learning. A significant positive association was also observed between LST and BTS proficiency ($r_s$ = 0.531, $p$ = 0.019), indicating a moderate association between the perceived importance of ICT proficiency for teaching and learning and BTS proficiency. However, the association between LST and CP proficiency did not reach statistical significance ($r_s$ = 0.395, $p$ = 0.094), indicating that the association between the perceived importance of ICT proficiency for teaching and learning and career progression towards proficiency in these applications was not significant, except for SA. The use of ICT as a teaching tool allows students to interact with various media, emphasizing information gathering, analysis, and organization (Çakici, 2016). ICT, which is also known as concept-based learning, is used in conjunction with multimedia computers. Albion et al. (2015) further stated that the expanding opportunities for ICT integration in all aspects of the school environment present teachers in the 21st century with new challenges.

Table 4. Spearman's Association Report

| Associations | | WPA | SA | PRE | CA | PVEA |
|---|---|---|---|---|---|---|
| LST | Association | 0.411 | .664** | 0.442 | .625** | .583** |
| | Sig. | 0.08 | 0.002 | 0.058 | 0.004 | 0.009 |
| CP | Association | 0.289 | 0.361 | 0.294 | 0.291 | 0.243 |
| | Sig. | 0.23 | 0.128 | 0.222 | 0.227 | 0.315 |

*Association is significant at the 0.05 level*

## CONCLUSIONS

The respondents were considered expert in using a spreadsheet application. However, they are only competent in terms of basic computer troubleshooting. This means that the respondents were more proficient in using spreadsheet applications and had the least knowledge of basic troubleshooting. In addition, the results indicated that the use of ICT applications is highly significant in the areas of learning, teaching, and career progression. The findings highlight the importance of integrating technological tools and resources into educational settings and professional development. Moreover, the results indicated that learning and teaching were significantly positively correlated with proficiency in using a variety of ICT applications, including word processing, spreadsheets, presentations, video conferencing, photo and video editing, and basic computer troubleshooting. Except for the spreadsheet application, the data showed that there was no connection between ICT proficiency in teaching and learning and career progression toward expertise in these applications.

## RECOMMENDATIONS AND IMPLICATIONS

In light of the results presented in this study, which indicate lower proficiency in troubleshooting, the researchers suggest implementing targeted workshops designed to augment teachers' troubleshooting skills. Such workshops can equip educators with the knowledge and competence required to identify and rectify technical issues that may arise when using ICT applications. By enhancing their troubleshooting capabilities, teachers can be better prepared to effectively overcome technical challenges.

Recognizing the paramount significance of integrating ICT into educational environments, researchers strongly advocate the continued integration of ICT applications and resources within the classroom context. By incorporating these technologies into the curriculum, PE teachers can progressively acclimate themselves to a diverse array of ICT-related applications. This not only fosters their familiarity with these tools, but also empowers educators to harness the potential of ICT to craft innovative and engaging teaching methodologies, thereby enriching their students' learning experiences. To further enhance the proficiency and competence of teachers in the effective utilization of ICT tools, the researchers propose the implementation of comprehensive training sessions and seminars. These educational programs should concentrate on a broad spectrum of ICT applications, including word processing, spreadsheet manipulation, presentation software, video conferencing platforms, and photo- and video-editing tools, among other pertinent ICT programs. By equipping educators with advanced skills in these areas, such training initiatives can significantly elevate their capabilities not only in ICT, but also in their professional growth and employability prospects.

In both academic and professional contexts, the ongoing assessment and evaluation of an individual's ICT skills play a pivotal role. Regular assessments serve as a valuable diagnostic tool, pinpointing specific areas that may require improvement and allowing for the customization of training programs to meet individual needs. Furthermore, soliciting feedback from respondents regarding their training or seminar experiences and their interactions with ICT applications can yield invaluable insights. This feedback can be leveraged to refine and adapt ICT-related interventions, ensuring their effectiveness and relevance in the evolving landscape of technology-enhanced education and professional development.

This study encourages future researchers to broaden the scope of their investigations to encompass a wider array of pertinent ICT applications and emerging technologies that could have implications for teaching, learning, and career advancement. Staying abreast of the latest technological advancements in the realm of education is imperative. Continuously monitoring these developments and evaluating their impact on educational enhancement should be integral to ongoing research efforts. By doing so, researchers can contribute to the dynamic and evolving landscape of educational technology, ensuring that their work remains relevant and influential in shaping the future of education

## ACKNOWLEDGEMENT

The authors would like to thank all the people who supported this study.

## FUNDING

The study did not receive funding from any institution.

## DECLARATIONS

### *Conflict of Interest*

The authors hereby declare that there is no conflict of interest associated with this study.

*Informed Consent*

All respondents in this study provided informed consent, which was seamlessly integrated into the data collection questionnaire. This ensured they were fully aware of the nature of the study and their participation in it.

*Ethics Approval*

No ethics approval was necessary for this study, but it followed Philippine Health Research Ethics Board (PHREB) guidelines and complied with local data privacy laws to protect the personal information of the respondents.

## Author's Biography

Kristine Joy D. Magallanes, Mark Brianne C. Carreon, Kristalyn C. Miclat, Niña Vina V. Salita, Gino A. Sumilhig, and Raymart Christopher C. Guevarra are preservice physical education teachers at Don Honorio Ventura State University.

John Paul P. Miranda is an associate professor at Don Honorio Ventura State University.